\documentclass[prl,twocolumn,floatfix,superscriptaddress,reprint]{revtex4}

\usepackage{amsmath,amssymb,amsfonts,xspace}
\usepackage{mathrsfs}
\usepackage{epsfig,supertabular,hyperref}

\DeclareMathOperator{\Li}{Li}

\newcommand{\be}{\begin{equation}}
\newcommand{\ee}{\end{equation}}
\newcommand{\beq}{\begin{eqnarray}}
\newcommand{\eeq}{\end{eqnarray}}
\newcommand{\bea}[2]{\be\label{#2}\begin{array}{#1}}
\newcommand{\eea}{\end{array}\ee}

\def\Tr{\,{\rm Tr}\, }

\def\Im{\,{\rm Im}\, }
\def\Re{\,{\rm Re}\, }

\def\({\left(}
\def\){\right)}
\def\[{\left[}
\def\]{\right]}
\def\p{\partial}

\def\11{1\!\! 1}

\def\eps{\varepsilon}

   \def\CI {{\cal I}}

   \def\CX {{\cal X}}
   \def\CY {{\cal Y}}

\def\bX{\bar X}
\def\bY{ \bar Y }

\def\bF{\bar F}

\def\bi{\bar {\i}}

\def\bt{\bar t}

\def\bv{\bar v}

\newcommand{\bZ}{\bar Z}

\newcommand{\de}{\mathrm{d}}

\newcommand{\I}{\mathrm{i}}

\newcommand{\cH}{\mathcal{H}}
\newcommand{\cI}{\mathcal{I}}

\newcommand{\half}{\frac{1}{2}}

\newcommand{\cM}{\mathcal{M}}

\newcommand{\cN}{\mathcal{N}}

\newcommand{\cX}{\mathcal{X}}

\newcommand{\cO}{\mathcal{O}}

\newcommand{\cU}{\mathcal{U}}

\newcommand{\cZ}{\mathcal{Z}}
\newcommand{\IR}{\mathbb{R}}
\newcommand{\IC}{\mathbb{C}}

\newcommand{\IP}{\mathbb{P}}

\def\Hij#1{H^{[#1]}}

\def\xii#1{\xi_{[#1]}}

\def\txii#1{{\tilde\xi}^{[#1]}}
\def\ai#1{{\alpha}^{[#1]}}

\def\ui#1{^{[#1]}}

\def\varpi{t}

\def\pa{\partial}

\newcommand{\kahler}{{K\"ahler}\xspace}
\newcommand{\hk}{{hyperk\"ahler}\xspace}
\newcommand{\qk}{{quaternion-K\"ahler}\xspace}

\def\bse{\begin{subequations}}
\def\ese{\end{subequations}}

\def\qli2{{\bf E}}

\def\varpihk{t}

\def\CY{\mathfrak{Y}}

\def\pzeta{\partial^{(t)}}
\def\bpzeta{\bar\partial^{(t)}}

\newcommand{\genpot}{\mu}

\def\In#1#2{\cI^{(#1)}_{#2}}

\begin{document}

\title{
\vskip -80pt
{\begin{normalsize}
\mbox{} \hfill LPTA/14-023, CERN-PH-TH-2014-100, arXiv:1406.2360v2 \\

\vskip 10pt
\end{normalsize}}

{\bf\Large
$\IR^3$ index for four-dimensional $\cN=2$ field theories
} }

\author{Sergei Alexandrov}
\email{salexand@univ-montp2.fr}
\affiliation{Laboratoire Charles Coulomb, CNRS UMR 5221,
Universit\'e Montpellier 2, F-34095 Montpellier, France}

\author{Gregory W. Moore}
\email{gmoore@physics.rutgers.edu}
\affiliation{NHETC and Department of Physics and Astronomy, Rutgers University, Piscataway, NJ 08855-0849, USA}

\author{Andrew Neitzke}
\email{neitzke@math.utexas.edu}
\affiliation{Department of Mathematics, University of Texas at Austin, Austin, TX 78712-1202, USA}

\author{Boris Pioline}
\email{boris.pioline@cern.ch}
\affiliation{CERN Dep PH-TH, 1211 Geneva 23, Switzerland\\
on leave from  CNRS, UMR 7589, LPTHE, F-75005, Paris, France \\
and Sorbonne Universit\'es, UPMC Univ. Paris 06, UMR 7589, LPTHE, F-75005, Paris, France
}

\begin{abstract}
\noindent In theories with $\cN=2$ supersymmetry on $\IR^{3,1}$,
BPS bound states can decay across walls
of marginal stability in the space of Coulomb branch parameters, leading to discontinuities in the BPS indices $\Omega(\gamma,u)$.
We consider a supersymmetric index $\cI$ which receives contributions from 1/2-BPS states, 
generalizing the familiar Witten index $\Tr (-1)^F  e^{-\beta H}$.
We expect $\cI$ to be smooth away from loci where massless particles appear, 
thanks to contributions from the continuum of multi-particle states.
Taking inspiration from a similar phenomenon
in the hypermultiplet moduli space of $\cN=2$ string vacua, we conjecture a formula
expressing $\cI$ in terms of
the BPS indices $\Omega(\gamma,u)$, which is continuous across the walls and exhibits 
the expected contributions from single particle states at large $\beta$.
This gives a universal prediction for the contributions of multi-particle states to the index $\cI$.  
This index is naturally a function on the moduli space after reduction on a circle, 
closely related to the canonical hyperk\"ahler metric and hyperholomorphic connection on this space.
\end{abstract}

\maketitle

It has been clear since the work of Seiberg and Witten \cite{Seiberg:1994rs} that
extended supersymmetry gives enough control over four-dimensional
quantum field theories to produce exact results on the \emph{dynamics} of
the theories, even when these theories are strongly interacting. Remarkably, such results
are deeply related to some of the most interesting questions in the mathematics of algebraic geometry and differential geometry.
As a significant example, the moduli space of a four-dimensional theory with $\cN=2$ supersymmetry on a circle is a hyperk\"ahler manifold
(a special class of manifolds satisfying Einstein's equations), whose metric encodes both instanton corrections
to gauge couplings and the spectrum of BPS states in the four-dimensional theory \cite{Gaiotto:2008cd}.
In this Letter, we reinforce this connection, and construct a canonical function on the afore-mentioned moduli space,
which on the one hand generates a solution to the self-dual Yang-Mills equations on this manifold,
and on the other hand, purportedly encodes interactions of BPS states in four dimensions.

\section{BPS indices and Witten index}

In four-dimensional field theories on $\IR^{3,1}$
with $\cN=2$ supersymmetry, the spectrum
of BPS states in general strongly depends on the value of the Coulomb branch parameters.
Part of this dependence can be removed by considering the BPS index
\be
\label{defOmega}
\Omega(\gamma,u)=-\tfrac12\Tr_{\cH_1(\gamma,u)}(-1)^{2J_3}(2J_3)^2,
\ee
where $\cH_1(\gamma,u)$ is the Hilbert space  of one-particle states with electromagnetic charge
$\gamma\in\Gamma$ in the Coulomb vacuum $u$, $J_3$ is a component of the rotation group along a fixed axis, and $(-1)^{2J_3}$
is the fermionic parity by virtue of the spin statistics theorem.
The BPS index $\Omega(\gamma,u)$, being sensitive only to short
multiplets saturating the BPS bound \cite{Witten:1978mh}, is a locally
constant, integer valued function of $u$, but it is discontinuous across certain
walls in moduli space, where some of the BPS bound states with charge $\gamma$ decay into
multi-particle BPS states \cite{Seiberg:1994rs,Ferrari:1996sv}.
The jump of $\Omega(\gamma,u)$ across the walls is governed by a universal wall-crossing
formula \cite{ks}, which can be derived by quantizing the configurational degrees of freedom of
multi-centered BPS states near the wall \cite{Denef:2007vg,Andriyash:2010qv,Manschot:2010qz}
(see e.g. \cite{Pioline:2011gf} for a  review).

The present work addresses another apparently protected quantity, the Witten index
\be
\label{defWittenZ}
\cI(\beta,u,C) =
-\tfrac12\Tr_{\!\cH(u)} (-1)^{2J_3}(2J_3)^2  \sigma\, e^{-\beta H-2\pi\I \langle \gamma, C \rangle},
\ee
where $\cH(u)$ is the full Hilbert space of the four-dimensional theory on $\IR^3$.
Here,
$\beta$ is the inverse temperature, conjugate to the Hamiltonian $H$, $C$
are chemical potentials conjugate to the electromagnetic charge $\gamma$,
and $\sigma$ is an operator on $\cH(u)$ acting by a sign $\sigma_{\gamma}$
in the sector with charge $\gamma$, such that
$ \sigma_\gamma \sigma_{\gamma'} = (-1)^{\langle \gamma,\gamma'\rangle} \sigma_{\gamma+\gamma'}$,
where $\langle \gamma, \gamma' \rangle$
is the usual Dirac-Schwinger-Zwanziger product ---
this sign is crucial in ensuring the consistency of self-dual field theories
\cite{AlvarezGaume:1986mi,AlvarezGaume:1987vm,Witten:1996hc,Freed:2000ta,Belov:2006jd}.
For simplicity, we restrict to theories without flavor charges.
The use of the canonical ensemble with respect to the electromagnetic charges is not essential,
but facilitates the geometric interpretation of the index.

Most importantly, unlike the well-studied case of  the index on $S^3$, the spectrum of the Hamiltonian on $\IR^3$
is gapless, due to massless gauge bosons and their superpartners, and
continuous, as it includes all multi-particle states
made out of the discrete states in $\bigoplus_{\gamma\in\Gamma} \cH_1(\gamma,u)$.
The contribution of the point spectrum to the index \eqref{defWittenZ}
is  controlled by the BPS indices  $\Omega(\gamma,u)$,
and is therefore discontinuous across walls of marginal stability.
Multi-particle states, on the other hand,
can also contribute despite the fact that they do not saturate the BPS bound, due to a possible
spectral asymmetry between bosonic and fermionic states \cite{Akhoury:1984pt}.
Our main assumption is that the Witten index \eqref{defWittenZ} is continuous across walls of marginal stability, 
as a result of cancellations between discontinuities from single and multi-particle state contributions. 
This assumption is physically reasonable, since the path integral defining \eqref{defWittenZ} 
suffers no phase transition across the wall. Under this assumption, we propose a formula
for expressing \eqref{defWittenZ} in terms of the BPS indices $\Omega(\gamma,u)$.

Our assumption is further supported by analogy with the case of framed BPS indices associated
to line defects in $\cN=2$  theories of class S.
These indices
are defined by a formula almost identical to \eqref{defWittenZ} (without the insertion of $(2J_3)^2$),
and are known to be smooth across BPS-walls  \cite{Gaiotto:2010be}. The Witten index \eqref{defWittenZ}
can be viewed as the  extension of the framed index to the case
of a trivial line defect.

Another class of examples where a Witten-type index is known to be a smooth function of the moduli
arises in $\cN=2$ supersymmetric massive theories in 1+1 dimensions: the BPS indices
$\Omega_{ab}$, which count
single-particle kinks interpolating between pairs $ab$ of supersymmetric
vacua, exhibit similar wall-crossing phenomena as in 3+1 dimensions \cite{Cecotti:1992rm},
while the CFIV index $\Tr_{ab} (-1)^F F e^{-\beta H}$ is continuous across the walls, as a result
of cancellations between single-particle and multi-particle contributions 
\cite{Cecotti:1992qh}\footnote{{In fact, the authors of \cite{Cecotti:1992qh} proposed a variant of \eqref{defWittenZ}
in 3+1 dimensions, investigated later by S. Sethi and C. Vafa (unpublished).}}.

Yet another way to support our assumption is the general expectation that the Witten index
controls quantum corrections to BPS-saturated couplings in the low-energy effective action.
In the case of four-dimensional $\cN=2$ theories, an appropriate  coupling is the metric on
the moduli space of the theory reduced on a spatial circle of radius $R$. In 2+1 dimensions,
Abelian gauge fields can be dualized into scalar fields, and the dynamics on the Coulomb branch can be formulated
as a non-linear sigma model. Its target space $\cM_3(R)$ is a torus fibration over
the Coulomb branch moduli space $\cM_4$ in 3+1 dimensions, with the torus fiber
parametrizing the holonomies $C$ of the electromagnetic gauge
fields around the circle \cite{Seiberg:1996nz}.
Supersymmetry requires the metric on $\cM_3(R)$ to be \hk (HK). In the limit $R\to\infty$
it is obtained from the special \kahler metric on $\cM_4$ via
the so-called `rigid $c$-map' procedure \cite{Cecotti:1989qn}.
For finite radius, however, the metric on $\cM_3(R)$ receives instanton corrections of
order $e^{-R}$ from BPS states in 3+1 dimensions, whose Euclideanized worldline winds around the
circle \cite{Seiberg:1996nz,Gaiotto:2008cd}
(a supersymmetric version of a mechanism first envisaged in \cite{Polyakov:1976fu}).
Although corrections to the metric components include an infinite series of multi-instanton corrections,
they are entirely controlled by the BPS indices $\Omega(\gamma,u)$ counting single-particle states.
Furthermore, it is manifest from the twistorial construction of $\cM_3(R)$ \cite{Gaiotto:2008cd}
that the quantum corrected metric is regular across walls of marginal stability, with multi-instanton
contributions on one side of the wall replacing
the one-instanton correction on the other side (alternatively, the smoothness of
the metric on $\cM_3(R)$ provides
a physical rationale for the wall-crossing formulae of \cite{ks}).

Since quantum corrections to the moduli space metric in theories with 8 supercharges are
generally saturated by 1/2-BPS contributions, it is natural to expect a connection between
the metric on $\cM_3(R)$ and the Witten index $\cI(\beta,u,C)$ for $\beta=2\pi R$.
The goal of this paper is to construct a natural function
on the family of  spaces $\cM_3(R)$, continuous across the walls,
which reproduces the expected contributions of single-particle states to the Witten index in the limit $R\to \infty$.
We conjecture that these two functions are equal, which allows us to predict
the contributions
of the continuum of multi-particle states to $\cI(\beta,u,C)$.

The clue for our construction comes from an analogous problem in superstring theory,
namely the vector multiplet moduli space $\widetilde{\cM}_3$ in type IIA/B  string vacua
of the form $\IR^3 \times S^1(R) \times \CY$ where $\CY$ is a  Calabi-Yau threefold.
After T-duality on the circle and its decompactification, the same moduli
space describes the hypermultiplet sector of the dual type IIB/A theory on
$\IR^4\times\CY$ \cite{Seiberg:1996ns}.
In contrast to the gauge theory case, $\widetilde{\cM}_3$ is a \qk (QK) manifold, where $R$ appears as one of the coordinates.
In the limit $R\to\infty$, the metric is obtained by the `local $c$-map' procedure from
the vector multiplet moduli space $\widetilde{\cM}_4$ in type IIA/B on $\IR^4 \times \CY$
\cite{Ferrara:1989ik}, whereas for finite radius it receives $\cO(e^{-R})$
corrections from four-dimensional BPS states winding around the circle (T-dual to D-instantons).
These instanton corrections can be incorporated through the twistor space construction \cite{Alexandrov:2008gh,Alexandrov:2009zh} 
(see \cite{Alexandrov:2011va,Alexandrov:2013yva} for reviews).
However, unlike the gauge theory set-up, the instanton series is divergent due to
the exponential growth of the BPS indices. Arguably, this is resolved by the existence of further gravitational (or NS5-brane)
instanton corrections of order $e^{-R^2}$ \cite{Pioline:2009ia}.
In the sector with zero NUT charge, which is insensitive to these additional instantons,
the twistorial construction of \cite{Alexandrov:2008gh,Alexandrov:2009zh} is formally isomorphic
to the gauge-theoretic one \cite{Gaiotto:2008cd}, specialized to the case of theories with a non-anomalous $U(1)_R$ symmetry,
described by a homogeneous prepotential $F(X)$ of degree 2.
This isomorphism was shown to be a particular instance
of a general correspondence between QK metrics with quaternionic $U(1)$ action and HK metrics
with $U(1)$ isometry rotating the complex structures \cite{Haydys,Alexandrov:2011ac,Hitchin:2013iga}
(the correspondence proceeds by lifting the $U(1)$ action on the QK
manifold to the Swann bundle and then taking the HK quotient).
Through this correspondence, the family of HK metrics $\cM_3(R)$ inherits a canonical function,
the moment map of the $U(1)_R$ action,
which is smooth as long as the metric on $\cM_3(R)$ is.
On the QK side it appears as the `contact potential', which relates the $O(2)$-twisted
canonical one-form to the holomorphic contact one-form  \cite{Alexandrov:2008nk,
Alexandrov:2008gh}.
As we shall see, a `generalized contact potential' can be defined
even when the prepotential $F$ is not homogeneous.
Geometrically, it can be understood as the ratio of two Hermitian metrics
on the canonical line bundle constructed in \cite{Neitzke:2011za}.

\section{A family of smooth functions on $\cM_3(R)$}

To define our candidate for the Witten index $\cI$,
let us first recall the twistorial construction of the HK metric on
$\cM_3(R)$ \cite{Gaiotto:2008cd} {(see \cite{Neitzke:2013tca} for a review)}.
The twistor space $\cZ=\IP_t\times \cM_3(R)$ carries a family
of  functions $\{\cX_\gamma(t)\}_{\gamma\in\Gamma}$,
holomorphic in complex structure $J(t)$, satisfying
the integral equations~\footnote{For brevity, we omit the dependence of $\Omega(\gamma,u)$
on the Coulomb branch parameters. Since the BPS indices are piecewise constant,
this dependence  does not spoil holomorphicity away from the walls.
}
\be
\frac{\cX_\gamma}{\cX_\gamma^{\text{sf}}}= \exp\!\!\left[\sum_{\gamma'}
\frac{\Omega(\gamma')}{4\pi\I}\left<\gamma, \gamma'\right>
\!\! \int_{\ell_{\gamma'}} \!\!\!\! \frac{\text{d}t'}{t'}
\frac{t+t'}{t-t'}\log\(1- \cX_{\gamma'}(t')\)\right] ,
\label{TBA-HK}
\ee
where $\ell_{\gamma'}$ are the BPS rays $\{ t' \in \mathbb{C}^{\times}:\  Z_{\gamma'}/t' \in \I\IR^-\}$ and
$\cX_\gamma^{\text{sf}}$ provide the boundary conditions at $R\to\infty$,
\be
\label{defXsf}
\cX^{\text{sf}}_\gamma=\sigma_{\gamma}\,
e^{-\pi \I R \left(t^{-1} Z_\gamma - t \bZ_\gamma \right)-2\pi \I\left<\gamma, C \right>}.
\ee
Here $Z_\gamma=\langle \gamma,X\rangle$ is the central charge and $X=(X^\Lambda,F_\Lambda)$ is the holomorphic symplectic section
on the special \kahler manifold $\cM_4$
with $F_\Lambda=\pa_{X^\Lambda}F$ in special coordinates.
In the limit $R\to\infty$, the system \eqref{TBA-HK} can be solved iteratively,
generating a sum of products of iterated integrals of $\cX_\gamma^{\text{sf}}$
interpreted as multi-instanton contributions.
Given such a solution, the triplet of \kahler forms on $\cM_3(R)$, which define the metric,
is read off from the $\cO(2)$-twisted holomorphic symplectic form on $\cZ$,
\be
\label{defomega}
\begin{split}
\omega =&\, \I t^{-1} \omega_+ + \omega_3 + \I t \omega_-
=
\frac{\epsilon^{ab}}{8\pi^2}\, \frac{\de \cX_{\gamma_a}}{\cX_{\gamma_a}} \wedge
\frac{\de \cX_{\gamma_b}}{\cX_{\gamma_b}},
\end{split}
\ee
where $\gamma_a$ is a basis of $\Gamma$, and $\epsilon^{ab}$ is the inverse of $\langle\gamma_a,\gamma_b\rangle$.

With these notations in place, fix any smooth function $F_\gamma(t,u,C)$
on $\Gamma\times \cZ$, linear in $\gamma$ and define
\be
\label{defPhi}
\Phi(R,u,C) = \sum_\gamma \Omega(\gamma)
\int_{\ell_\gamma} \frac{\text{d}t}{t}\,  F_\gamma\, \log\(1-\cX_\gamma\) .
\ee
We claim that  $\Phi$ is
a smooth function on $\cM_3(R)$, provided
the BPS indices $\Omega(\gamma)$ jump across walls of marginal stability
according to the standard wall-crossing formula \cite{ks}. Indeed, on a
wall $W(\gamma_1,\gamma_2)$, where the central charges $Z_{\gamma_1}$, $Z_{\gamma_2}$
associated to two primitive charge vectors become aligned in the complex plane,
the BPS rays $\ell_{m\gamma_1+n\gamma_2}$ with $m,n\geq 0$ all
coalesce into one ray $\ell$, across which the potential discontinuity is given by
\be
\label{discPhi}
\begin{split}
\Delta\Phi= &\,
\int_{\ell} \frac{\text{d}t}{t}  \sum_\gamma F_\gamma\, \Bigl[ \Omega^+(\gamma)
\log\(1-\cX^+_\gamma \)
\Bigr.
\\
&\, \Bigl.
-\Omega^-(\gamma)
\log\(1-\cX^-_\gamma \) \Bigr] ,
\end{split}
\ee
where $\Omega^\pm(\gamma)$ and $\cX^\pm_\gamma$ are the BPS indices and
solutions of the corresponding equations \eqref{TBA-HK} on either side of the wall.
Now, recall  that the
semi-classical limit of the motivic version of the wall-crossing formula implies
the functional identity  \cite{Alexandrov:2011ac}
\be
\label{wcdilog}
\sum_\gamma \Omega^+(\gamma)\, L_{\sigma_{\gamma}}( \cX^+_\gamma ) =
\sum_\gamma \Omega^-(\gamma)\, L_{\sigma_{\gamma}}( \cX^-_\gamma )
\ee
where $L_\eps(z)$ is a variant of the Rogers dilogarithm,
\be
\label{defrogersv}
L_\eps(z) \equiv {\rm Li}_2(z)+\frac12\, \log (\eps^{-1} z) \log(1-z) .
\ee
The invariance of \eqref{wcdilog} under monodromies
$M_\gamma: \cX_{\gamma'} \mapsto e^{2\pi\I\langle \gamma,\gamma'\rangle} \cX_{\gamma'}$ leads to
the $\Gamma$-valued identity
\be
\label{wcddilog}
\sum_\gamma \gamma\[\Omega^+(\gamma) \log\(1-\cX^+_\gamma \) - \Omega^-(\gamma) \log\(1-\cX^-_\gamma \) \]=0 .
\ee
The vanishing of the discontinuity \eqref{discPhi} then follows from \eqref{wcddilog}
and from the linearity of $F_\gamma$ with respect to $\gamma$.

\section{A candidate for the Witten index}

Having constructed a family of smooth functions on $\cM_3(R)$, we now aim for
one that may plausibly be identified with
the Witten index \eqref{defWittenZ}.
For HK manifolds $\cM_3(R)$ related to QK manifolds  $\widetilde{\cM}_3$
by the QK/HK correspondence, a natural candidate is the contact potential
on $\widetilde{\cM}_3$ \cite{Alexandrov:2008nk,Alexandrov:2008gh},
which translates on the HK side into
\be
\label{defZc}
\cI=\!\!
 \frac{ R}{16\I\pi^2}
\sum_\gamma \Omega(\gamma)
\int_{\ell_\gamma} \frac{\text{d}t}{t}\left( t^{-1}  Z_\gamma  -t\bar Z_\gamma \right) \log\(1-\cX_\gamma(t)\).
\ee
This function is a member of the family \eqref{defPhi} with
$F_\gamma(t) \propto t^{-1} \,Z_\gamma\, - t \bar Z_{\gamma}$, so it is
smooth across walls of marginal stability. Its reality follows from the reality
property $\overline{\vphantom{t^A}\CX_{\gamma}(-1/\bar t)}=\CX_{-\gamma}(t)$
and the CPT relation $\Omega(-\gamma) = \Omega(\gamma)$.

In order to assess whether \eqref{defZc} qualifies
to represent the Witten index, let us compute the formal multi-instanton expansion of $\cI$, by
substituting the iterated solution of \eqref{TBA-HK} into \eqref{defZc}. Up to second order,
$\cI=\sum\limits_{\gamma} \In{1}{\gamma} + \sum\limits_{\gamma,\gamma'} \In{2}{\gamma,\gamma'}+\dots$ with
\beq
\In{1}{\gamma}&=&  \frac{R}{4\pi^2}\,
\sigma_{\gamma} \,\overline{\Omega}(\gamma)\,
|Z_\gamma|K_1(2\pi  R |Z_\gamma|)\, e^{-2\pi\I  \langle \gamma,C \rangle},
\nonumber\\
\In{2}{\gamma,\gamma'}&=& -\frac{ R}{64\pi^3}\,
\overline{\Omega}(\gamma) \, \overline{\Omega}(\gamma') \, \, \langle \gamma,\gamma'\rangle
\int_{\ell_\gamma} \frac{\text{d}t}{t} \,
\int_{\ell_\gamma'} \frac{\text{d}t'}{t'}\frac{t+t'}{t-t'}
\nonumber\\
&&\times\,
\left( t^{-1}  Z_\gamma  -t\bar Z_\gamma \right)
\cX_\gamma^{\rm sf}(t)  \cX_{\gamma'}^{\rm sf}(t')\, ,
\label{2instZ}
\eeq
{where $\overline{\Omega}(\gamma)=\sum_{d|\gamma} \tfrac{1}{d^2}
\Omega(\gamma/d)$ denotes the rational index.}
Remarkably, {for primitive charge vector $\gamma$} the one-instanton contribution $\In{1}{\gamma}$ agrees with the contribution
of a single-particle, relativistic BPS state of charge $\gamma$ and mass  $M=|Z_\gamma|$
to the Witten index  \eqref{defWittenZ}. To see this, we use a Schwinger time parametrization
to linearize the relativistic Hamiltonian $H=\sqrt{-\Delta+M^2}$, and introduce
a non-zero chemical potential $\theta$ conjugate to $J_3$ and periodic
boundary conditions $\psi(z)=\psi(z+L)$ along the $z$ axis, with $L\gg 1/M$, to regulate infrared divergences.
Denoting by $\chi_{\rm spin}(\theta)$ the $SU(2)$ character for
the spin degrees of freedom,  we have
\beq
\Tr e^{-2\pi R H+\I\theta J_3} \!  &=  & R \int_0^{\infty} \!\frac{\de t}{t^{3/2}}
\Tr e^{-\pi \frac{R^2}{t}- \pi (-\Delta+M^2) t+\I\theta J_3}
\nonumber\\
=  & R & \!\!\!\! \int_0^{\infty}\! \frac{\de t}{t^{3/2}}
\frac{L}{2\pi\sqrt{t}}\, \frac{\chi_{\rm spin}(\theta)}{4\sin^2(\theta/2)}\, e^{-\pi \frac{R^2}{t}-\pi M^2 t}
\nonumber\\
&= & \frac{L}{2\pi}  \frac{\chi_{\rm spin}(\theta)}{4\sin^2(\theta/2)} \,2M\, K_1(2\pi M R)\, .
\eeq
For a BPS multiplet of spin $j$, the spin character is
\be
\chi_{\rm spin}(\theta) = \(2+2\cos\frac{\theta}{2}\) \frac{\sin\[ (j+\tfrac12) \theta\]}{\sin(\theta/2)}\, ,
\ee
corresponding to a BPS index $\Omega(\gamma)=2\pa_{\theta}^2 \chi_{\rm spin}(\theta) \vert_{\theta=2\pi} = (-1)^{2j} (2j+1)$.
Comparing with the first line in \eqref{2instZ} we find
\be
\label{Z1vsTr}
\In{1}{\gamma}= 2R\!\lim\limits_{\substack{\theta\to 2\pi\\ L\to\infty}} \pa_{\theta}^2 \left[ \frac{\sin^2(\theta/2)}{\pi L}
\,
\Tr\! \(\sigma\,e^{-2\pi R H+\I\theta J_3-2\pi\I \langle \gamma,C \rangle}\)
\right]\! .
\ee
The factor $\sin^2\tfrac{\theta}{2}/(2\pi L)$
can be understood as dividing by the regularized volume of $\IR^3$.

Based on this agreement, and smoothness across walls of marginal stability, we conjecture
that \eqref{defZc} in fact computes the Witten index \eqref{defWittenZ}, with the specific
prescription given in \eqref{Z1vsTr} for regulating infrared divergences.
If true, this implies that the two-instanton term $\In{2}{\gamma,\gamma'}$
in \eqref{2instZ} should be identified with the contribution of the continuum of two-particle states,
 and similarly for higher $\In{n}{}$'s.

\section{Geometric nature of $\cI$}

While the Ansatz \eqref{defZc} was suggested by the QK/HK correspondence, we now wish
to elucidate its nature from the viewpoint of the HK geometry of $\cM_3(R)$, without
assuming that the prepotential $F(X)$ is homogeneous.
For this purpose, we need
to recall an additional construct on the HK space $\cM_3(R)$,
namely the canonical hyperholomorphic line bundle $\mathscr{L}$ introduced
in  \cite{Neitzke:2011za}, generalizing the one afforded by the QK/HK correspondence in the homogeneous
case \cite{Haydys,Alexandrov:2011ac}. In twistorial terms, $\mathscr{L}$ descends
from a line bundle $\mathscr{L}_{\cZ}$ on $\cZ$ determined by the local holomorphic section
$\hat\Upsilon=\Upsilon_{\rm sf} \Upsilon_{\rm inst}$,
given in unitary gauge by
\beq
\label{defUpsilon}
\Upsilon_{\rm sf}\!&=&
e^{-\I\pi \[ \frac{R^2}{4} \(\frac{2F-X^\Lambda F_\Lambda}{t^2} +t^2 (2\bF-\bX^\Lambda \bF_\Lambda)\)
+ \frac{R}{2}\langle t^{-1}X-t \bX,C \rangle \]},
\nonumber\\
\Upsilon_{\rm inst}\! &=& \exp\Biggl[ \frac{1}{8\pi^2} \sum_\gamma \Omega(\gamma)
\int_{\ell_\gamma} \frac{\text{d}t'}{t'}\, \frac{t+t'}{t-t'} \biggl( L_{\sigma_{\gamma}}( \cX_\gamma(t'))
\biggl.\Biggr.
\nonumber\\
&& \Biggl.\biggl. -
\frac{1}{2}\,\log \frac{ \cX^{\rm sf}_\gamma(t)}{ \cX^{\rm sf}_\gamma(t')}
\log\( 1-  \cX_\gamma(t') \)\biggr) \Biggr].
\eeq
The line bundle $\mathscr{L}_{\cZ}$ is equipped with a family of connections,
represented by the one-forms
\be
\label{deflambdaA}
\hat\lambda(t) = \frac{\eps^{ab}}{8\pi^2} \, \log\cX_{\gamma_a}\, \de\log\cX_{\gamma_b} -\frac{1}{2\pi\I}\, \de\log\hat\Upsilon ,
\ee
whose curvature is given by the holomorphic symplectic form \eqref{defomega}.
We define the ``generalized contact potential'' as the contraction
of $\hat\lambda(t)$ with the vector field $\kappa=t\pa_t$ on $\cZ= \IP_t\times \cM$ generating
the $\IC^\times$ action on the first factor and leaving the second factor invariant,
\be
\hat\mu(t) = -\I\( \iota_{\kappa} \hat\lambda(t)\) .
\label{defmu}
\ee
This is a function on $\cZ$, meromorphic on each fiber. {A computation shows that}, upon adding a suitable
``classical" term,
\be
\label{defItot}
\cI_{\rm tot}\equiv \cI- \frac{R^2}{2}\,\Im(\bar X^\Lambda F_\Lambda)
\ee
the completed
(purported) Witten index $\cI_{\rm tot}$
is equal to the constant term in the Laurent expansion of $\hat\mu(t)$ around $t=0$, or equivalently,
$t=\infty$.

By the twistor correspondence \cite{Ward:1977ta},
the complex line bundle $\mathscr{L}_{\cZ}$ on $\cZ$ descends to a line bundle with
a hyperholomorphic connection on $\cM_3(R)$. In the same unitary trivialization
as above, it can be represented by the one-form $\hat\lambda(t) + \check\lambda(t)$
where
\beq
&&\hspace{-0.6cm}
\check\lambda(t)=\frac{R}{8\pi^2}\Biggl[ 2\pi\I \, \epsilon^{ab} \( t^{-1}Z_{\gamma_a}  -t\bZ_{\gamma_a} \)
\de\log\cX_{\gamma_b}
\Biggr.
\label{deflambdaB}\\
&&\hspace{-0.6cm}
\Biggl.
+\sum_{\gamma}
\Omega(\gamma)
\!\! \int\limits_{\ell_{\gamma}}  \frac{\text{d}t'}{t'} \Big( \frac{ Z_\gamma}{t'}  -t'\bZ_\gamma \Big)
\frac{t' \pa^{(0)}+t \bar\pa^{(0)}}{t'-t}\log\(1- \cX_{\gamma}(t')\)\Biggr]
\nonumber
\eeq
is a (1,0)-form in complex structure $J(t)$, smooth across BPS rays, such that
$\hat\lambda(t) + \check\lambda(t)$ is independent of $t$ \footnote{In \eqref{deflambdaB},
$\pa^{(t)}$ denotes the Dolbeault operator in complex structure $J(t)$. The first term in
\eqref{deflambdaB} is manifestly of type $(1,0)_t$. So is the second one, due  to the fact that
the operator $t' \pa^{(0)}+t \bar\pa^{(0)}$ maps $J(t')$-holomorphic functions
into $(1,0)_t$ forms.}. The curvature $\mathscr{F} =  \de(\hat\lambda+\check\lambda)$
is  then hyperholomorphic, i.e. of type (1,1) in any complex structure $J(t)$ \cite{Neitzke:2011za,Alexandrov:2011ac},
and is given by
\be
\label{Fhyper}
\mathscr{F}
= \frac{\I}{\pi}\, \pzeta\bpzeta \Re \log \hat\Upsilon.
\ee

Our second claim is that the constant term
$\check\lambda_0$ in the Laurent expansion of $\check\lambda(t)$ around $t=0$ (or equivalently $t=\infty$)
satisfies
\be
\check\lambda_0 = \I \bigl( \pa^{(0)} - \bar\pa^{(0)} \bigr) \cI_{\rm tot}.
\ee
This implies that the completed Witten index
determines the difference between the hyperholomorphic curvature
$\mathscr{F}$ and the \kahler form for the instanton corrected HK metric on
$\cM_3(R)$ in complex structure $J(0)$ via
\be
\label{Kdiff}
\begin{split}
\omega_3 - \mathscr{F} =&\, 2\I\p^{(0)}\bar\p^{(0)}  \cI_{\rm tot}.
\end{split}
\ee
This equation generalizes the well-known statement in the homogeneous case that
the moment map $\mu$ of the $U(1)_R$ action with respect to $\omega_3$
(which coincides with the contact potential on the QK side) is such that
$\omega_3-2\I \pa^{(0)}\bar\pa^{(0)} \mu$ is a hyperholomorphic
two-form \cite{Haydys,Hitchin:2013iga}.
One way to express \eqref{Kdiff} geometrically is to say that $\cI_{\rm tot}$ is equal
to the logarithm of the ratio of two different Hermitian metrics on the line bundle $\mathscr{L}_\cZ$, whose curvatures are
equal to the \kahler form $\omega_3$ and the hyperholomorphic curvature $\mathscr{F}$, respectively.

Another interesting consequence of
\eqref{Kdiff} is that $\cI_{\rm tot}$ is a quasi-harmonic function on $\cM_3(R)$,
\be
\label{LaplacianI}
\Delta  \cI_{\rm tot} = - 4r ,
\ee
where $\Delta$ is the Laplace-Beltrami operator for the HK metric on $\cM_3(R)$
and the right-hand side is minus the real dimension of $\cM_3(R)$. The condition
\eqref{LaplacianI} is reminiscent of the second order differential equations which typically constrain BPS saturated amplitudes.

In the Appendix we show that the geometrical objects introduced in this section can be
defined for a general class of HK manifolds, and obtain a generalization of Eq. \eqref{Kdiff}
valid in any complex structure $J(t)$.

\section{Discussion}

In this article we conjectured a formula \eqref{defZc} for the generalized Witten index \eqref{defWittenZ}
in four-dimensional $\cN=2$ gauge theories. The formula is manifestly smooth
across walls of marginal stability, and correctly reproduces the expected
BPS bound states contributions. The evidence for this conjecture is admittedly weak, since
within the class \eqref{defPhi} of smooth functions on the Coulomb branch $\cM_3(R)$ in
three dimensions, one could easily find other functions  which would differ only at higher
order in the multi-particle expansion. Our proposal however is distinguished by the fact that
a completed version \eqref{defItot} of $\cI$
is related to the \kahler form and hyperholomorphic curvature on $\cM_3(R)$ via \eqref{Kdiff},
in accordance with the general slogan that corrections to the moduli space metric in
theories with 8 supercharges are  saturated by 1/2-BPS contributions. It would be
interesting to derive the ``classical term" in \eqref{defItot} from the contribution
of massless gauge bosons and supersymmetric partners, and extend our construction
to gauge theories with massive flavors~\footnote{Massless flavors do not pose any difficulty,
but non-zero masses disturb the integrality of the periods of the curvature of $\mathscr{L}$.}.
We note that the function $\cI$ has already appeared in the context of
the analogy of the system \eqref{TBA-HK} with TBA equations
\cite{Gaiotto:2008cd}, where it is identified with
the free energy of the corresponding integrable system \cite{Alexandrov:2010pp},
and in the context of minimal surfaces in $AdS_5$ \cite{Alday:2009yn,Alday:2009dv}.

If correct, our conjecture predicts that multi-particle state contributions to the Witten index are universal
functions of the BPS indices $\Omega(\gamma)$ associated to the constituents.
The predicted
contribution of the continuum of two-particle states can be found in \eqref{2instZ},
while higher orders can be easily obtained by combining \eqref{defZc}
with the iterated solution to the TBA-like system \eqref{TBA-HK}.
It is a challenge to check these predictions from a direct computation of the difference of densities
of bosonic and fermionic  states of a system of $n$ dyons. While the result near a wall of marginal
stability can actually be deduced by analyzing the non-relativistic electron-monopole
system \cite{Pioline:2015wza},
the result \eqref{2instZ} should hold throughout moduli space, where the constituents are relativistic.

Note also that our conjecture naturally extends to the case of $\cN=2$ string vacua,
where the formula \eqref{defZc} computes instanton corrections to the contact potential
on the QK moduli space $\widetilde\cM_3$ generated by multi-dyonic BPS black holes.
Therefore, another check
would be to reproduce the smooth, duality invariant
partition function for two-centered D4-D2-D0 black holes
constructed in \cite{Manschot:2009ia}, extending the arguments in \cite{Alexandrov:2012au}
beyond the one-instanton level.

As we have mentioned, the generalized Witten index  \eqref{defWittenZ} may be viewed as the analog of the framed BPS
index for a trivial line defect. One possible way to derive \eqref{defWittenZ}
would then be to study the fusion of two line defects whose OPE contain the trivial line defect.
This analogy also suggests the existence of a refined Witten index,
which would arise in the fusion of framed protected spin characters. It is natural to conjecture that
this refined index might be related to the CFIV index {of the two-dimensional theory
obtained by placing the four-dimensional theory on an $\Omega$-background with 
$\epsilon_1\neq 0,\epsilon_2=0$ \footnote{{We thank D. Gaiotto and N. Nekrasov for suggesting this interpretation.}
}}.

Finally, our conjecture -- if true -- could reveal interesting and nontrivial
information on BPS spectra which is not easily accessible by other
means. For example, consider a theory of class S where the ultraviolet
curve $C$ is a compact Riemann surface with negative curvature.
In this case $\cI_{\rm tot}$ is just the moment map for the natural $U(1)$
action on Hitchin data and hence proportional
to the $L^2$ norm-square of the Higgs field  \cite{Hitchin:2013iga}.
The expression \eqref{defZc} is highly nontrivial already in the $A_1$ case. In this case one may be able
to give a systematic large $R$ expansion of the norm-square of the
Higgs field by solving the classical sinh-Gordon theory on $C$. Using the
parametrization of \cite[Eq. (13.14)]{Gaiotto:2009hg}, it is easy to show that,
on a real slice of moduli space one needs to expand
\begin{equation}
\CI_{\rm tot} = \frac{\I R^2}{4} \int_C  \lambda \bar \lambda \cosh (2h)
\end{equation}
at large $R$ for   solutions to the sinh-Gordon equation
\begin{equation}
\p \bar\p h - 2R^2 \lambda\bar\lambda \sinh(2h)=0
\end{equation}
with boundary condition $h \sim - \half \log \vert z -z_a \vert+ \cdots $
at the first order zeros $z=z_a$ of the quadratic differential $\lambda^2$. We hope to return to this
problem in a future publication.

\section*{Acknowledgments}

\noindent  We are grateful to S. Cecotti, D. Gaiotto, N. Nekrasov, A. Sen, S. Sethi, J. Troost and C. Vafa   for useful discussions.
SA and BP are grateful to Daniel Persson, Stefan Vandoren and Frank Saueressig for past collaboration
which led to some of the constructions underlying the present discussion. The work of   GM is supported by the DOE under grant
DOE-SC0010008 to Rutgers,   and NSF Focused Research Group
award DMS-1160461.
The work of AN is supported by NSF award DMS-1151693 and Focused Research Group award DMS-1160461.

\section{Appendix}

In this section we elaborate
on the geometric origin of the function $\cI_{\rm tot}$
and establish \eqref{Kdiff} as a special case of a more general formula derived from the twistorial construction
of HK manifolds developed in \cite{Alexandrov:2008ds}.

For a HK manifold $\cM$, let us choose an atlas $\cup \cU_i$ covering
the twistor space $\cZ=\IP_t\times \cM$ and a set of Darboux coordinate
systems $\Xi_a^{[i]}=(\xii{i}^\Lambda,\txii{i}_\Lambda)$, $\Lambda=1\dots n$,
regular everywhere on $\cU_i$ except for
$\xii{+}^\Lambda$ and $\xii{-}^\Lambda$, which are allowed to have simple poles at
two special points
$t=0$ and $t=\infty$.
The real structure is assumed to map $\Xi_a^{[i]}(t)$ to $\overline{\Xi_a^{[\bi]}(-1/\bt)}$,
where $\cU_{\bi}$ is the patch opposite to $\cU_i$ under the antipodal map.
The holomorphic symplectic structure on $\cZ$ is specified by
a set of local holomorphic functions $\Hij{ij}(\xii{i},\txii{j},t)$
which generate symplectomorphisms on overlaps $\cU_i \cap \cU_j$ of two patches,
\be
\xii{j}^\Lambda=\xii{i}^\Lambda-\p_{\txii{j}_\Lambda }\Hij{ij},
\qquad
\txii{j}_\Lambda =  \txii{i}_\Lambda + \p_{\xii{i}^\Lambda } \Hij{ij} ,
\label{glucon}
\ee
subject to the obvious reality and cocycle conditions on triple overlaps $\cU_i \cap \cU_j\cap \cU_k$.
The HK manifold $\cM$ parametrizes the space of solutions $\Xi_a^{[i]}(t)$
to the gluing conditions \eqref{glucon}. Substituting  $\Xi_a^{[i]}(t)$ into
$\omega^{[i]}= \de\txii{i}_\Lambda \wedge\de\xii{i}^\Lambda$ and equating with \eqref{defomega}
gives access to the triplet of \kahler forms, hence to the HK metric.

The set of transition functions $\Hij{ij}$ naturally defines a holomorphic
affine bundle $\mathscr{L}_{\cZ}$ on $\cZ$, whose sections satisfy the following gluing conditions
\be
\ai{j}=\ai{i}+\Hij{ij}-\xii{i}^\Lambda\p_{\xii{i}^\Lambda } \Hij{ij} .
\label{holsec}
\ee
The consistency of \eqref{holsec} on triple overlaps is ensured by the cocycle condition on $\Hij{ij}$.
The affine bundle $\mathscr{L}_\cZ$ carries a connection
which gives rise to a holomorphic one-form
\be
\label{deflami}
\lambda^{[i]}\equiv -\de\ai{i}-\xii{i}^\Lambda\de\txii{i}_\Lambda.
\ee
We use it to define the ``generalized contact potential'' as in \eqref{defmu}
\be
\label{defZi}
\genpot^{[i]}(\varpi)\equiv -\I\(\iota_{\kappa}\lambda^{[i]}\) .
\ee
Using the system of integral equations equivalent to the gluing conditions \eqref{glucon}, \eqref{holsec}
(see \cite[Eq.(2,11)]{Alexandrov:2011va}), one obtains~\footnote{In this formula the operator $t\pa_t$
should not be confused with the vector field $\kappa$ appearing in \eqref{defZi}.
Here it just acts on the last argument of the function $\Hij{ij}(\xii{i},\txii{j},t)$.}
\be
\begin{split}
\genpot^{[i]}(\varpi)=&\, \frac{1}{4\pi}\sum_j\oint_{C_j}\frac{\de\varpi'}{\varpi'}\biggl[
\(\varpi'^{-1} Y^{\Lambda}-\varpi' \bY^{\Lambda} \)\p_{\xi^\Lambda } \Hij{ij}
\biggr.
\\
&\, \biggl.
+\frac{\varpi'+\varpi}{\varpi'-\varpi}\,\varpi'\p_{\varpi'} \Hij{ij}\biggr],
\end{split}
\label{gencontpot}
\ee
where $C_j$ surround the patches and $Y^\Lambda$ ($\bY^\Lambda$) is the residue of $\xii{\pm}^\Lambda$ at $t=0$ ($\infty$).
Moreover, using again the integral expression for $\ai{i}$, one can prove that
\be
K^{[i]}(t)= 2\Re \genpot^{[i]}(t)- 2\,\frac{1-|t|^2}{1+|t|^2}\,\Im \ai{i}
\label{Kdiffgen}
\ee
provides a \kahler potential for the HK metric in complex structure $J(t)$.
In the presence of a
$U(1)_R$ isometric action, the transition functions $\Hij{ij}$
are independent of $t$, so the last term in \eqref{gencontpot} disappears and $\genpot^{[i]}$
becomes independent of $t$ as well (and hence on the patch index $i$).
Under the QK/HK correspondence, $\genpot$
is identified with the contact potential on the dual QK manifold \cite{Alexandrov:2011ac}.

Let us now explain the relation between this construction and the one given in the main text as well as
why \eqref{Kdiffgen} generalizes \eqref{Kdiff} to arbitrary complex structure.
For this purpose, we need to specify the atlas and the transitions
functions $\Hij{ij}$ relevant for the HK manifold $\cM_3(R)$.
The atlas consists of the patches $\cU_\pm$ around the north and south poles, and the patches
$\cU_{\gamma}$ lying in between two consecutive BPS rays, the one on the right being $\ell_\gamma$ (see \cite[Fig. 4.2]{Alexandrov:2011va}).
Then we take
\be
\Hij{+\gamma}=\frac{R^2}{4\varpihk^2} F\(\frac{2\varpihk\xi}{R}\),
\quad
\Hij{-\gamma}=\frac{(R\varpihk)^2}4 \bF\(-\frac{2\xi}{\varpihk R}\)
\ee
as transition functions from $\cU_\pm$ to $\cU_{\gamma}$,
and
\be
\begin{split}
H_\gamma =&\,  G_{\gamma} (\cX_\gamma) - \frac12\, p^\Lambda q_\Lambda[G'_{\gamma}(\cX_\gamma)]^2,
\\
G_\gamma(\cX)  =&\, \frac{\Omega(\gamma)}{(2\pi)^2}\Li_2(\cX),
\end{split}
\ee
for transition functions across $\ell_\gamma$. Here
$\gamma=(p^\Lambda,q_\Lambda)$ and $\cX_\gamma=\sigma_{\gamma}\,e^{-2\pi \I  \left<\gamma, \Xi\right>}$.
The generalized contact potential \eqref{gencontpot} is then given in
any patch $\cU_{\gamma}$ by
\be
\label{defZt}
\genpot^{[\gamma]}(t) =  \cI_{\rm tot}
+\frac{R^2}{4\I}\( t^{-2}f-\varpi^2\bar f\)
+\frac{R}{2\I}\( t^{-1}f_\Lambda v^\Lambda+\varpi\bar f_\Lambda\bv^\Lambda\),
\ee
where
$f=2F-X^\Lambda F_\Lambda$ and $v^\Lambda$ is the constant term in the Laurent expansion of $\xii{+}^\Lambda$.
Thus, like $\hat\mu$ in \eqref{defmu}, the constant term in the Laurent expansion
$\genpot^{[\gamma]}$ generates the (purported, completed) Witten index.

To understand the relation between $\genpot^{[\gamma]}$ and $\hat\mu$, and between
$\alpha^{[\gamma]}$ and $\hat\Upsilon$,  let us introduce a variant of $\alpha$,
\be
\hat\alpha^{[i]} =\ai{i}+\frac12\sum_j \oint_{C_j} \frac{\de \varpi'}{2 \pi \I \varpi'} \,
\frac{\varpi'+\varpi}{\varpi'-\varpi}\frac{1-t'^2}{1+t'^2}\, t'\p_{t'}\Hij{ij},
\label{alpha-dif}
\ee
which satisfies (compare with \eqref{holsec})
\be
\hat\alpha^{[j]}=\hat\alpha^{[i]}+\Hij{ij}-\xii{i}^\Lambda\p_{\xii{i}^\Lambda } \Hij{ij} +\frac{t^{-1}-t}{t^{-1}+t}\, t\p_t\Hij{ij}.
\label{holsec-mod}
\ee
Similarly we define $\hat\lambda\ui{i}$ as in \eqref{deflami} with $\ai{i}\to \hat\alpha\ui{i}$.
One can check that $\hat\Upsilon$ in \eqref{defUpsilon} and $\hat\lambda$ in \eqref{deflambdaA}
are related to hatted quantities in the patches $\cU_{\gamma}$ via
\be
\label{hatUp}
\hat\Upsilon=e^{\pi\I\(2\hat\alpha^{[\gamma]}+\xii{\gamma}^\Lambda\txii{\gamma}_\Lambda\)},
\qquad
\hat\lambda = \hat\lambda^{[\gamma]} .
\ee
In particular, the relation $\lambda^{[\gamma]}-\hat\lambda^{[\gamma]}=
\de \bigl(\hat\alpha^{[\gamma]}-\alpha^{[\gamma]}\bigr)$ explains why
the constant terms in $\genpot^{[\gamma]}$ and $\hat\mu$ coincide.
It is worth noting that for a homogeneous prepotential, $\hat\alpha\ui{i}$ coincides with $\ai{i}$,
and the construction of \cite{Alexandrov:2011ac} is recovered. If $F$ is not homogeneous,
the unhatted counterpart of \eqref{hatUp} formally produces another hyperholomorphic curvature
on the universal cover of $\cM_3(R)$, albeit one which is inconsistent
with symplectic invariance and periodicity under integral shifts of $C$.

Finally, let us  derive \eqref{Kdiff} from the general property \eqref{Kdiffgen}.
In the patch around the north pole, one obtains
\beq
\genpot^{[+]}(0)&=&\cI_{\rm tot}+\frac{R^2}{4\I}\,\bX^\Lambda f_\Lambda
+\frac{1}{2\I}\, F_{\Lambda\Sigma\Xi}X^\Lambda v^\Sigma v^\Xi.
\label{muzero}
\\
\ai{+}(0)&=&-\frac{\I}{2\pi}\, \lim_{t\to 0}\[  \log\hat\Upsilon
+\text{hol.}\, \]
\label{alphazero}\\
&+&
\frac{\I}{2}\Im\[R^2 f +\frac{R^2}{2}\,\bX^\Lambda f_\Lambda
+ F_{\Lambda\Sigma\Xi}X^\Lambda v^\Sigma v^\Xi\],
\nonumber
\eeq
where $\text{hol.}$ denotes a holomorphic function needed to cancel the pole of $\log\hat\Upsilon$ at $t=0$.
Substituting \eqref{muzero} and \eqref{alphazero} into \eqref{Kdiffgen}
one finds
\be
K(0)=2\,\cI_{\rm tot} + \frac{1}{\pi} \Re  \lim_{t\to 0}\[ \log\hat\Upsilon
+\text{hol.} \right]-R^2\Im f.
\ee
Eq. \eqref{Kdiffgen} then follows by
applying the operator $\I \pa^{(0)}\bar\pa^{(0)}$
and taking into account \eqref{Fhyper}.


\end{document}